\documentclass[AMA,STIX1COL]{WileyNJD-v2}
\articletype{Research article}%
\received{XX.XX.XXXX}
\revised{YY.YY.YYYY}
\accepted{ZZ.ZZ.ZZZZ}

\usepackage{graphicx}

\usepackage{booktabs}

\jname{Research Synthesis Methods}
\articletype{RESEARCH ARTICLE}
\jyear{2025}

\begin{document}

\providecommand{\expect}{\mathrm{E}}             
\providecommand{\var}{\mathrm{Var}}              
\providecommand{\normaldistn}{\mathrm{Normal}}   
\providecommand{\unifdistn}{\mathrm{Uniform}}    
\providecommand{\halfnormaldistn}{\mathrm{half\mbox{-}Normal}}
\providecommand{\expodistn}{\mathrm{Exp}}
\providecommand{\halftdistn}{\mathrm{half}\mbox{-}\mathrm{Student}\mbox{-}t}
\providecommand{\halflogidistn}{\mathrm{half}\mbox{-}\mathrm{logistic}}
\providecommand{\lomaxdistn}{\mathrm{Lomax}}
\providecommand{\elir}{\mathrm{ESS}_{\mathrm{ELIR}}}  
\providecommand{\mup}{\mu_{\mathrm{p}}}        
\providecommand{\sigmap}{\sigma_{\mathrm{p}}}  



\title{Meta-analytic-predictive priors based on a single study}

\author[1]{Christian R\"{o}ver}
\author[1,2,3]{Tim Friede}

\authormark{Christian R\"{o}ver, Tim Friede}

\address[1]{\orgdiv{Department of Medical Statistics}, \orgname{University Medical Center G\"{o}ttingen}, \orgaddress{\state{G\"{o}ttingen}, \country{Germany}}.\email{christian.roever@med.uni-goettingen.de}}
\address[2]{\orgname{DZHK (German Center for Cardiovascular Research)}, \orgdiv{partner site Lower Saxony}, \orgaddress{\state{G\"{o}ttingen}, \country{Germany}}}
\address[3]{\orgname{DZKJ (German Center for Child and Adolescent Health)}, \orgaddress{\state{G\"{o}ttingen}, \country{Germany}}}



\abstract[Summary]{Meta-analytic-predictive (MAP) priors have been proposed as a generic approach to deriving informative prior distributions, where external empirical data are processed to learn about certain parameter distributions. The use of MAP priors is also closely related to shrinkage estimation (also sometimes referred to as \emph{dynamic borrowing}). A potentially odd situation arises when the external data consist only of \emph{a single study}. Conceptually this is not a problem, it only implies that certain prior assumptions gain in importance and need to be specified with particular care. We outline this important, not uncommon special case and demonstrate its implementation and interpretation based on the normal-normal hierarchical model. The approach is illustrated using example applications in clinical medicine.}


\maketitle



\section*{Highlights}
\paragraph{What is already known:}
\begin{itemize}
  \item shrinkage estimation may be used to effectively and robustly borrow information between related data sources
  \item shrinkage estimation may alternatively be motivated via a meta-analytic-predictive (MAP) approach
\end{itemize}

\paragraph{What is new:}
\begin{itemize}
  \item a MAP approach remains sensible down to the extreme case of only a single study
  \item the MAP prior's usual features are retained, in addition there are connections to power prior and bias allowance approaches
\end{itemize}

\paragraph{Potential impact for RSM readers outside the authors’ field:}
\begin{itemize}
  \item MAP priors are useful for constructing empirically motivated priors based on  external/historical data
  \item MAP priors may serve as an additional motivation for related approaches (bias allowance models, power priors)
  \item practical application is straightforward using existing software packages
\end{itemize}

\section{Introduction}\label{sec:intro}
  The potential of clinical research is commonly limited by data sparsity issues; such problems particularly arise in the context of rare diseases, where the number of potential study subjects is small, or in pediatric indications, where ethical considerations may limit the recruitment of patients. The large variety of rare diseases still means that a sizeable proportion of the population is affected by rare diseases, posing a considerable economic burden. Even in more common indications, data sparsity problems may arise, for example, when the focus is on smaller sub-populations, or when novel treatments or standards of care emerge. In any of these cases, the careful consideration of all potentially relevant evidence available is essential \cite{GagneEtAl2014,TudurSmithEtAl2014,GamaloSiebersEtAl2017}.
  When evidence from a single experiment, such as a clinical trial, is not sufficiently conclusive on its own, it may sometimes help to view the data in the context of related instances (similar experiments) in order to yield more confident conclusions.
  This idea is explicitly implemented in \emph{shrinkage estimation}, where a hierarchical model is set up accounting for estimation uncertainly at the study level as well as for variability (and similarity) between studies \cite{MorrisLysy2012,GelmanHill};
  models of this kind are commonly also used in the context of meta-analysis \cite{Fleiss1993,Roever2020}.
  The borrowing-of-information taking place between the study of primary interest and the external data may be viewed in terms of the overarching joint model as a \emph{meta-analytic-combined (MAC)} approach, or, equivalently, by formulating the \emph{meta-analytic predictive (MAP)} prior that explicates the information contributed by the external data to the shrinkage estimate \cite{SchmidliEtAl2014}.
  A particular special case is given when a single (``target'') study is supported by a single (``source'') study; such situations are not uncommon, and shrinkage estimation here has proven useful \cite{RoeverFriede2020,RoeverFriede2021,LesaffreEtAl2024}.
  Application of a hierarchical model makes it behave \emph{dynamically} in the sense that more or less information is borrowed, depending on the apparent similarity of target and source data \cite{RoeverFriede2020}.
  When considering this case in terms of the implied MAP~prior, the ``meta-analysis'' involved here is based on a single study, which may appear somewhat counterintuitive at first.
  It is this perceived contradiction that we aim to address here; while a \emph{meta-analysis} is commonly thought of as involving larger amounts of data, we will see that a hierarchical model may essentially be fit also to a single data point, and sensible predictions may be derived.
  Such a smallest-possible meta-analysis does not pose a conceptual problem, and there is no reason to abandon the general concept even if the amount of historical data drops below a couple of studies. It
  only implies that, due to the particular sparsity of data, prior specification within the model receives special importance, a problem which, however, is common in meta-analysis of few studies in general \cite{RoeverEtAl2021}, and which analogously applies for alternative (and closely related) borrowing methods, such as power priors or bias allowance models \cite{LesaffreEtAl2024,WeltonEtAl}.
  Since others appear to have struggled with or shied away from the idea of a single-study meta-analysis where in fact it may have been a viable option \cite{IglesiasEtAl2018,HarariEtAl2023}, it seems worthwhile to investigate this special case a bit closer.
  Closer inspection of this particular case then also highlights how the properties of this MAP prior materialize, as well as its close connection to bias allowance and power prior approaches.

  The remainder of this article is structured as follows; in Section~\ref{sec:shrink} the normal-normal hierarchical model (NNHM) is introduced, the meta-analysis model which then is the basis for shrinkage estimation between a pair of studies, and for the MAP~prior based on a single study.
  The ideas will be illustrated in two practical examples in the following Section~\ref{sec:map}. 
  Section~\ref{sec:alport} discusses an application in paediatric Alport syndrome that was originally formulated in terms of a shrinkage estimation problem. 
  Section~\ref{sec:cardio} introduces a trial design application in cardiology, where information from a similar past study is designated for consideration in the eventual analysis via an informative MAP~prior. Section~\ref{sec:discussion} then closes with a brief discussion.


\section{Shrinkage estimation using two studies}\label{sec:shrink}
  \subsection{The normal-normal hierarchical model (NNHM)}\label{sec:nnhm}
    The most common model for random-effects meta-analysis is given by the \emph{normal-normal hierarchical model (NNHM)}. It implements sampling error as well as between-study heterogeneity using normal distributions. The data are given in terms of $k$~estimates~$y_i$ and associated standard errors~$s_i$ ($i=1,\ldots,k$). Each individual study aims to quantify a parameter~$\theta_i$, so that
    \begin{equation}
      y_i|\theta_i,s_i \;\sim\;\normaldistn(\theta_i, s_i^2)\mbox{.}
    \end{equation}
    The underlying parameters~$\theta_i$ are not necessarily identical for all studies, instead some amount of \emph{(between-study) heterogeneity} is allowed for, expressed as
    \begin{equation}\label{eqn:nnhm}
      \theta_i|\mu,\tau \;\sim\; \normaldistn(\mu,\tau^2) \mbox{.}
    \end{equation}

    Often the overall mean~$\mu$ is the aim of the analysis, while sometimes the \emph{study-specific parameters}~$\theta_i$ are also of interest \cite{WandelNeuenschwanderRoeverFriede2017,RoeverFriede2020}. The heterogeneity, while important, usually remains a nuisance parameter.
    In the context of ``shrinkage estimation'' of the~$\theta_i$, an interesting aspect is that the problem may be motivated in two ways; classically, one may think of shrinkage estimation as a joint analysis of all~($k$) estimates, which also returns estimates of any $\theta_i$~parameter along the way; this is also denoted as the \emph{meta-analytic-combined (MAC)} approach.
    The problem, however, may also be factored into the evidence stemming from the $i$th study alone, as well as the information provided by the remaining ($k-1$) estimates. Shrinkage estimation then may be interpreted as the analysis of the $i$th study, based on a prior distribution that results as the predictive distribution derived from a meta-analysis of the other ($k-1$)~studies; this prior is denoted as the \emph{meta-analytic-predictive (MAP) prior}. Both MAC and MAP approaches are equivalent and yield identical shrinkage estimates \cite{SchmidliEtAl2014}.
  
    In the following, we will focus on the special case of only two studies ($k=2$). For the shrinkage estimate  ($\theta_2$), this implies a MAP prior that is based on a single study (i.e., the data provided through~$y_1$ and~$s_1$).    
    While this may appear odd at first, the idea readily applies also in this special case, as will be demonstrated in the following. 
    Analysis may generally be performed based on informative or uninformative priors for~$\mu$, while a proper, informative prior is required for~$\tau$ \cite{Roever2020,RoeverEtAl2021}.
    The case of only $k=2$ studies
    is also closely connected to the related concepts of \emph{power prior} \cite{RoeverFriede2020} or \emph{bias allowance models} \cite{Pocock1976,WeltonEtAl}.

  \subsection{Uniform prior for the overall mean effect~($\mu$)}
    Priors for the overall mean parameter~($\mu$) in the NNHM may be specified as informative or as uninformative. For (more or less informative) priors, normal distributions are an obvious choice, also since these lead to analytically simple inference. Quite commonly, effect priors however are chosen as uninformative and (improper) uniform, not least due to certain analogies to frequentist meta-analysis procedures \cite{Roever2020}.
    In case of an (improper, non-informative) uniform prior for~$\mu$, certain expressions turn out particularly simple, which also is why we will focus on this particular, yet insightful, common and practically relevant case in the following.
    As the (improper) uniform prior constitutes the limiting case of an increasingly uninformative effect prior, the following considerations may also be viewed as relating to the limiting behaviour for increasingly uninformative priors (e.g., for normal priors when their variance approaches infinity).
    The uniform effect prior leads to a normal conditional posterior for the overall mean effect~$\mu$, with moments given by
    \begin{equation}\label{eqn:CondPostMom}
      \expect[\mu|y_1, s_1, \tau] \;=\; y_1 \quad \mbox{and} \quad
      \var(\mu|y_1, s_1, \tau)    \;=\; s_1^2 + \tau^2 \mbox{,}
    \end{equation}
    and a marginal heterogeneity likelihood that is constant (independent of~$\tau$), so that the heterogeneity's posterior equals its prior \cite{Roever2020}.
    This seems reasonable, since (as long as the overall mean effect prior is uniform) a single observation~$y_1$ does not provide information on the heterogeneity~$\tau$.

  \subsection{The MAP prior for the effect in a new study}
    The MAP prior results as the posterior predictive distribution
    for a ``new'', second study's (study-specific) effect~$\theta_2$ given the data from the first study ($y_1$, $s_1$).
    In the NNHM framework, the \emph{conditional predictive distribution} again is normal with mean
    \begin{equation}\label{eqn:CondPostPredMean}
      \expect[\theta_2|y_1, s_1, \tau] \;=\; \expect[\mu|y_1, s_1, \tau] \;=\; y_1
    \end{equation}
    and variance
    \begin{equation}\label{eqn:CondPostPredVar}
      \var(\theta_2|y_1, s_1, \tau)
      \;=\; \var(\mu|y_1, s_1, \tau) + \tau^2 
      \;=\; s_1^2 + 2\tau^2
    \end{equation}
    (see also (\ref{eqn:CondPostMom}) and~(\ref{eqn:nnhm}), and the more detailed derivation in Appendix~\ref{sec:PostPredAppendix}).
    These expressions make sense in the present context: We know~$\theta_1$ with accuracy given by the standard error~$s_1$, and we know that the difference between~$\theta_1$ and~$\theta_2$ is normally distributed with variance~$2\tau^2$, so that the (conditional) variance expression results as a corresponding sum \cite{RoeverFriede2020}.
    As pointed out above, information on heterogeneity~($\tau$) so far is based on the prior \emph{only};
    the heterogeneity in this context generally requires a proper, informative prior (since $k<3$) \cite{Roever2020,RoeverEtAl2021}.

    The eventual (marginal) predictive distribution (marginalized over the distribution of~$\tau$) hence results as a normal \emph{scale mixture} \cite{LeeMcLachlan2016,Lindsay} with fixed mean~(\ref{eqn:CondPostPredMean}) and with variance as given in~(\ref{eqn:CondPostPredVar}), where $\tau$~is distributed according to the specified prior.
    In particular, one may think of the MAP prior as the first study's point estimate~$p(\theta_1|y_1,s_1)$ convolved with the prior predictive distribution~$p(\theta_2|\theta_1,\tau)$ and then marginalized over~$\tau$.
    The MAP prior is symmetric around~$y_1$, and its (marginal) variance 
    results from~(\ref{eqn:CondPostPredVar}) as
    \begin{equation}\label{eqn:mapVar}
        \var(\theta_2|y_1, s_1) \;=\; \expect[s_1^2+2\tau^2] \;=\; s_1^2 + 2\,\expect[\tau^2] \mbox{,}
    \end{equation}
    where the expectation $\expect[\tau^2]$ depends on the assigned heterogeneity prior. For a range of common prior specifications, this expectation may be derived analytically; Table~\ref{tab:tauSquared} in Appendix~\ref{sec:MapVarAppendix} lists some popular cases.
    From this expression, one can see that the relative magnitudes of~$s_1^2$ and~$\expect[\tau^2]$ determine whether the resulting MAP prior's variance is dominated by estimation uncertainty (regarding~$\theta_1$) or anticipated heterogeneity~($\tau$).
    These two variance components may in fact also be considered to reflect so-called ``type~A'' and ``type~B'' uncertainties relating to measurement uncertainty and background knowledge, respectively, which together sum up to form the \emph{combined standard uncertainty}~$u_c$ (the square root of~(\ref{eqn:mapVar})) \cite{Kirkup2002,VanDerBlesEtAl2019}.

    Since the MAP prior results as a normal scale mixture, it is generally heavier-tailed than a normal distribution, which has implications for the resulting operating characteristics. A heavy-tailed (MAP-) prior means that in combination with a (``shorter-tailed'') normal likelihood, the likelihood will dominate in case of a prior-data conflict \cite{OHaganPericchi2012}.
    Such robustness properties have in fact been noted and demonstrated in the meta-analysis context, as these lead to a \emph{dynamic} borrowing behaviour \cite{RoeverFriede2020}.
    The MAP priors' tail behaviour will also be illustrated for some examples below (Figures~\ref{fig:predCdfs} and~\ref{fig:predLogDens}).

    Another way to quantify the precision of a prior distribution is by relating it to a number of observations that would in a certain sense convey an equivalent amount of information.
    In the following, we will use two approaches to this effect. Firstly, a prior may be assessed in terms of a corresponding absolute \emph{effective sample size (ESS)} (here: number of patients), this will be done by quoting  ESSs based on the \emph{expected local-information-ratio} ($\elir$). This measure is based on the prior density's curvature and it ensures \emph{predictive consistency}, that is, on expectation, the posterior's ESS will be the sum of the prior's $\elir$ \emph{plus} the actual sample size \cite{NeuenschwanderEtAl2020}.
    Secondly, the added information from including the prior in a specific analysis may be expressed in terms of the (relative) \emph{gain in effective sample size} \cite{RoeverFriede2020}. This is based on comparing the relative width of a confidence interval \emph{with} and \emph{without} considering the informative prior, and then determining by what factor the sample size would have needed to be increased to yield the same precision gain (see also Appendix~\ref{sec:EssGainAppendix}).

  \subsection{The bias allowance model connection}
    In the 2-study case, there is a one-to-one correspondence between the NNHM and a simple \emph{bias allowance model} \cite{WeltonEtAl}; instead of the NNHM assumption~(\ref{eqn:nnhm})
    in combination with a uniform prior for the overall mean effect~$\mu$
    and heterogeneity prior~$p_\star(\tau)$ as in Section~\ref{sec:nnhm}, one may specify
    \begin{eqnarray}
      \theta_2|\alpha,\beta &=& \alpha \\
      \theta_1|\alpha,\beta &\sim& \normaldistn(\alpha, \, \beta^2)
    \end{eqnarray}    
    with prior~$p(\beta)=\frac{1}{\sqrt{2}}\,p_{\star}\bigl(\frac{\beta}{\sqrt{2}}\bigr)$ for the standard deviation~$\beta$ \cite{RoeverFriede2020}.
    This ``reference model'' is different in that one estimate (the reference, or ``target''~$y_2$) directly relates to~$\alpha$, while the other one~(the ``source''~$y_1$) is associated with an additional offset to account for potential bias.
    The \emph{shrinkage estimates} of~$\theta_i$, however, can be shown to be identical in both models as long as a uniform prior for the overall mean effect~($\mu$) is used \cite{RoeverFriede2020}.
    The reference model may be considered a variation of \emph{Pocock's bias model} or, more generally, a \emph{bias allowance model} \cite{NeuenschwanderSchmidli2020,Pocock1976,WeltonEtAl}.
    The $\beta$~parameter, which only differs from~$\tau$ by a scaling factor of~$\sqrt{2}$, may also help motivating a heterogeneity prior, as it directly relates to the expected difference between~$\theta_2$ and~$\theta_1$, without reference to a common overall mean~$\mu$.

  \subsection{The power prior connection}\label{sec:powerPriorConnection}
    There is also is a connection to a so-called \emph{power prior}, which has been proposed as an approach for deliberate down-weighting of prior information. 
    It is intended for a prior distribution that itself results as a posterior, and the power prior results from applying an exponent~$a_0$ (with $0 \leq a_0 \leq 1$) to its likelihood contribution \cite{IbrahimChen2000,NeuenschwanderSchmidli2020}.
    When conditioning on a fixed $\tau$~value, the (conditional) MAP prior
    is normal with moments given in~(\ref{eqn:CondPostPredMean}) and~(\ref{eqn:CondPostPredVar}); in particular, note that $\tau^2$ acts additively on the ``plain'' variance~($s_1^2$).
    In this context, a power prior with fixed exponent~$a_0$ on the other hand would correspond to a $\normaldistn\Bigl(y_1,\,\frac{s_1^2}{a_0}\Bigr)$ distribution, where the (inverse) exponent acts multiplicatively on the variance.
    Both MAP and power prior then are identical if~$a_0=\bigl(2\frac{\tau^2}{s_1^2}+1\bigr)^{-1}$ \cite{ChenIbrahim2006,RoeverFriede2020,PawelEtAl2024}.
    It is interesting to note that the relationship between~$a_0$ and~$\tau$ here depends on the the ratio~$\frac{\tau}{s_1}$; Pawel \emph{et~al.} (2024) \cite{PawelEtAl2024} point out that the exponent~$a_0$ directly relates to the ``relative'' heterogeneity as expressed though the popular $I^2$~statistic \cite{HigginsThompson2002}, which in this case simply equals $I^2=\frac{\tau^2}{\tau^2+s_1^2}$. The exponent may then directly be expressed as a function of the corresponding~$I^2$ as $a_0=\frac{1-I^2}{1+I^2}$.
    While a prior probability distribution for~$\tau$ is readily motivated with reference to the effect scale of~$y_i$ and~$\theta_i$ \cite{RoeverEtAl2021}, specification of a fixed $\alpha_0$~value remains tricky, and a prior distribution may be even harder to motivate, as it would implicitly relate to the $I^2$~scale.
    On the other hand, through the above functional correspondence, any prior for the heterogeneity~$\tau$ immediately implies a corresponding distribution for the exponent~$a_0$; an example is illustrated in Appendix~\ref{sec:PowerPriorExponent}.


\section{Two practical applications}\label{sec:map}
    
  \subsection{Pediatric Alport example}\label{sec:alport}
  \subsubsection{Application}
    Gross \emph{et~al.} (2020) \cite{GrossEtAl2020} performed a randomized controlled trial (RCT) in Alport syndrome to investigate the effects of ramipril, an angiotensin-converting enzyme inhibitor (ACEi). 
    Recruitment of participants to the RCT was hampered by the rare and pediatric nature of the disease, and so the analysis of the RCT had been planned with the inclusion of observational data from an open-label arm and a natural disease cohort \cite{GrossEtAl2012}.
    \emph{Time to disease progression} was a co-primary endpoint, and from the observational data, a hazard ratio (HR) of~0.53 [0.22, 1.29] was estimated based on 70~patients. Only 20~patients entered into the RCT, and a HR of 0.51 [0.12, 2.20] was estimated. The data are also summarized in Table~\ref{tab:alport}.
    \begin{table}[ht]
      \centering
      \caption{\label{tab:alport}Alport example data from Gross \emph{et~al.} (2020) \cite{GrossEtAl2020}}
      \begin{tabular}{clcccc}
        \toprule
        &          &       &             & \multicolumn{2}{c}{log-HR} \\
        \cmidrule(lr){5-6}
        $i$ & estimate & $n_i$ & hazard ratio (HR) & $y_i$    & $s_i$ \\
        \midrule
        1   & observational & 70    & 0.53 [0.22, 1.29] & $-0.635$ & $0.451$ \\
        2   & RCT      & 20    & 0.51 [0.12, 2.20] & $-0.673$ & $0.742$ \\
        \bottomrule
      \end{tabular}
    \end{table}

    The analysis was then performed by jointly considering both log-HR estimates in a NNHM, anticipating a reasonable amount of heterogeneity between them (expressed through a $\halfnormaldistn(0.5)$~prior for~$\tau$), and deriving a shrinkage estimate for the RCT effect~($\theta_2$). The resulting estimate then was substantially more precise than if the RCT data were considered in isolation; the HR was estimated at 0.52 [0.19, 1.39] \cite{GrossEtAl2020}.

    To make the flow of information transparent, we may now derive the corresponding MAP prior reflecting the information contributed by the observational data.
  \begin{figure}
    \centering
    \makebox{\includegraphics[width=0.5\linewidth]{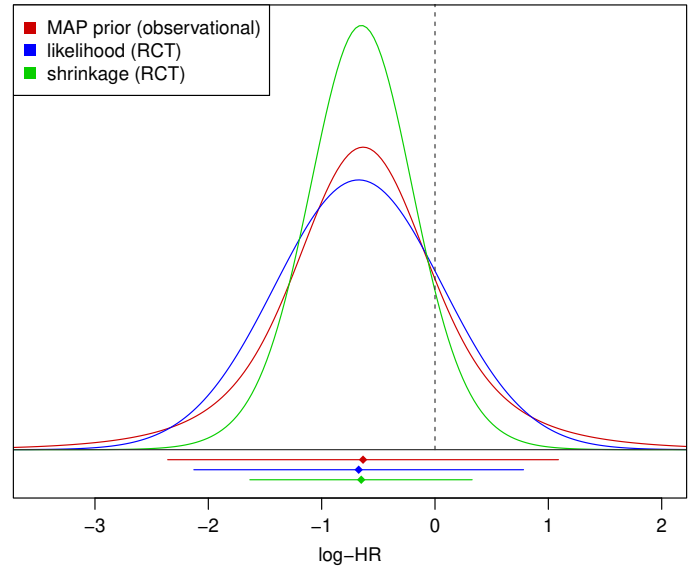}}
    \caption{\label{fig:alport}Illustration of MAP-prior, likelihood and (shrinkage-) posterior for the Alport example discussed in Section~\ref{sec:alport} \cite{GrossEtAl2020}. The horizontal lines at the bottom indicate point estimates and corresponding 95\% intervals} 
  \end{figure}
    Figure~\ref{fig:alport} illustrates MAP-prior, likelihood and posterior (shrinkage estimate) for the Alport example.
    The MAP prior here has a mean of $y_i=-0.63$ and a variance of $s_1^2 + 2\,\expect[\tau^2] = 0.45^2 + 2\times 0.5^2 = 0.84^2$.
    While the observational sample size was 70~patients,
    the MAP prior's effective sample size ($\elir$) is at only 26~patients (i.e., 37\% of originally 70~actual patients) \cite{NeuenschwanderEtAl2020}.
    The eventual shrinkage interval is only 67\% as wide as the original, implying a substantial ``effective gain in sample size'' \cite{RoeverFriede2020}; such a precision increase would otherwise have required more than doubling the sample size (by the addition of 24~extra patients). So in this case the absolute ($\elir$) estimate matches well the observed precision gain.

  \subsubsection{Variations of the MAP prior}\label{sec:variations}
    The heterogeneity ($\tau$) prior was specified as a $\halfnormaldistn(0.5)$ distribution, which is a reasonably conservative choice for endpoints such as HRs,
    as it covers ``reasonable'' and up to ``fairly high'' levels of heterogeneity ($\tau \leq 1$) and leaves a small prior probability for ``fairly extreme'' amounts ($\tau>1$)
    \cite{FriedeRoeverWandelNeuenschwander2017a,RoeverEtAl2021}.
    Since conclusions heavily depend on the heterogeneity prior settings, it may however be interesting to investigate the effects of a range of reasonable alternative specifications; in particular, we will consider different prior scales and different distribution families.
    Among the various assumptions implemented in the analysis, a ``too optimistic'' heterogeneity prior (favouring small heterogenity) might yield results inappropriately close to a common-effect analysis, while overly ``pessimistic'' or ``conservative'' assumptions may on the other hand may eventually lead to very little borrowing of information.
    
    Assuming that~$s_1=0.451$ (as in the present example, see Table~\ref{tab:alport}), we can illustrate the resulting MAP prior when varying the heterogeneity prior scale.
    \begin{figure}
      \centering
      \makebox{\includegraphics[width=0.5\linewidth]{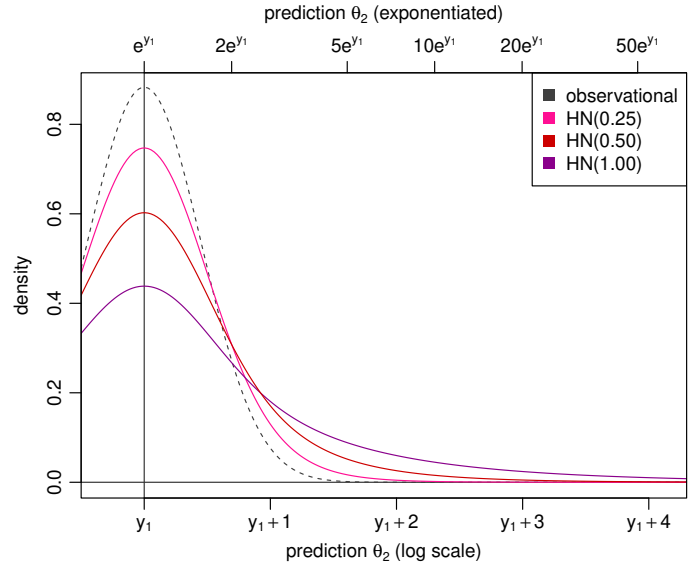}}
      \caption{\label{fig:scaleVariations}Illustration of the resulting  MAP-prior for varying heterogeneity prior scales. The dashed line indicates the likelihood of the observational data alone for comparison}
    \end{figure}
    Figure~\ref{fig:scaleVariations} shows the likelihood of the observational estimate along with the corresponding MAP priors for half-normal heterogeneity priors with scales~0.25, 0.50 and 1.00. Increasing the heterogeneity prior scale yields a MAP prior that becomes increasingly wider than the plain likelihood alone.
    In the present case, the effect scale was a logarithmic~HR, and on the logarithmic scale, the original MAP prior (based on a $\halfnormaldistn(0.5)$ heterogeneity prior) covers a range of $y_1 \pm 1.72$ with 95\% probability. On the exponentiated scale (see also the top axis in Figure~\ref{fig:scaleVariations}), a difference of 1.72 in log-HR would correspond to a 5.6-fold larger HR\@.
    If one switched to a $\halfnormaldistn(0.25)$ or $\halfnormaldistn(1.0)$ prior instead, the range would change to $y_1\pm 1.13$ or $y_1\pm 3.18$ instead, corresponding to multiplicative factors of~3.1 or~24.0, respectively.

    Half-normal distributions are a common and obvious choice as heterogeneity priors, possible reasons may be familiarity and availability, as well as a ``flat'' shape near the origin and a rather short tail \cite{RoeverEtAl2021}.
    Variations of the distribution family commonly do not alter conclusions dramatically as long as they cover a similar range, as manifested e.g. in a common prior median.

\begin{figure}
      \centering
      \makebox{\includegraphics[width=0.5\linewidth]{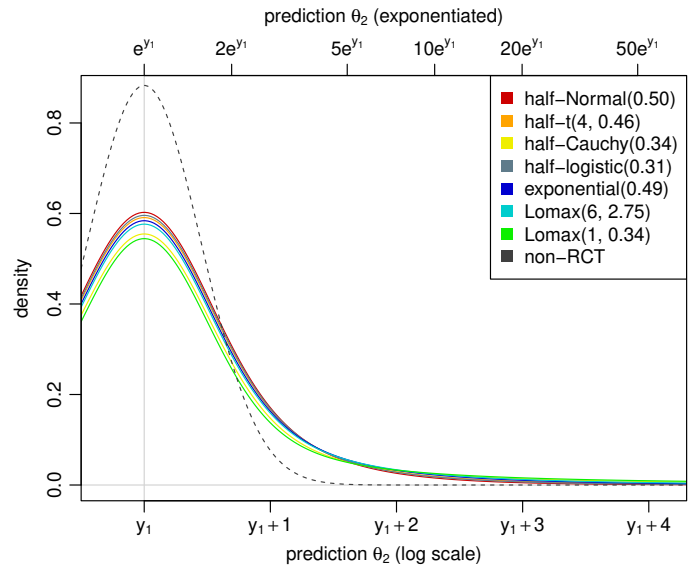}}
      \caption{\label{fig:predDensities}Illustration of MAP-prior's dependence on the heterogeneity prior distribution family. The different heterogeneity priors shown here all share the same prior median} 
    \end{figure}

\begin{figure}
      \centering
      \makebox{\includegraphics[width=0.5\linewidth]{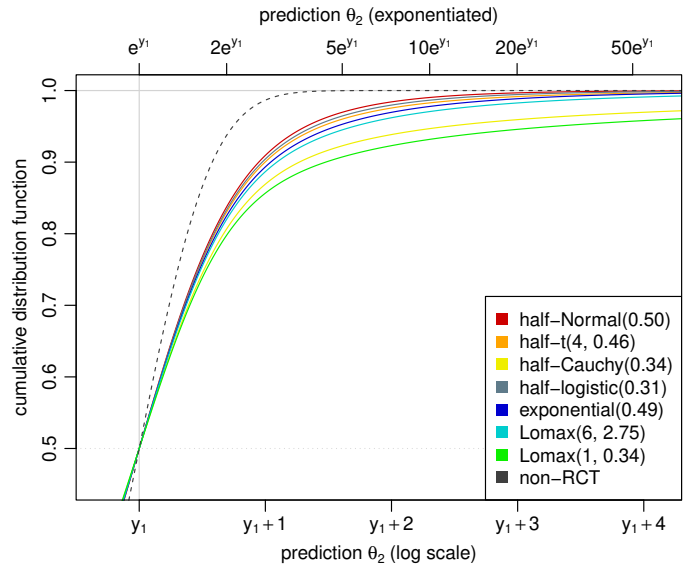}}
      \caption{\label{fig:predCdfs}The MAP-priors' cumulative distribution functions corresponding to the densities shown in Figure~\ref{fig:predLogDens}} 
    \end{figure}

\begin{figure}
      \centering
      \makebox{\includegraphics[width=0.5\linewidth]{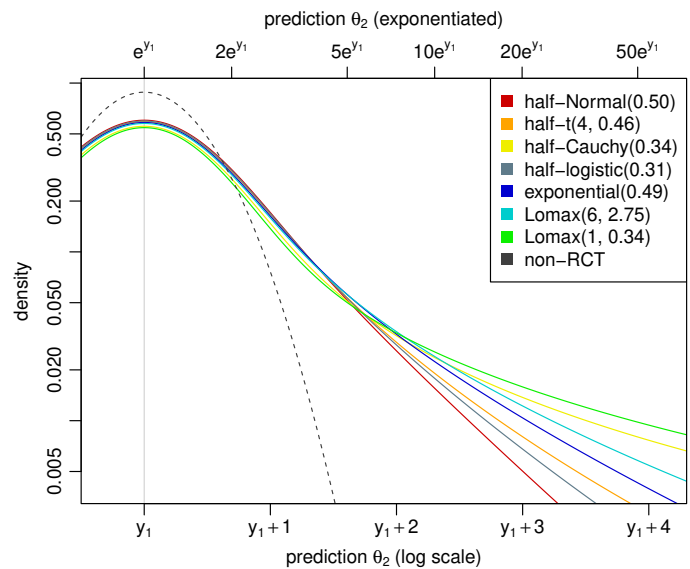}}
      \caption{\label{fig:predLogDens}The MAP-priors' densities on a logarithmic scale (see also Figure~\ref{fig:predLogDens}).
      Note that the likelihood for the observational data alone follows a parabola shape here, while the corresponding MAP~priors are clearly much heavier-tailed} 
    \end{figure}

    MAP priors corresponding to alternative specifications to a $\halfnormaldistn(0.5)$ for the heterogeneity are illustrated in Figure~\ref{fig:predDensities}. A range of distribution families are used, with their scale parameters specified such that all correspond to a common prior median for~$\tau$ (of~$0.34$).
    These different heterogeneity prior families are also shown in  Appendix~\ref{sec:PriorAppendix}. 
    The resulting MAP~prior densities themselves are hard to distinguish. Differences are more noticeable when focusing on the tail behaviour, e.g., considering cumulative distribution functions (as shown in Figure~\ref{fig:predCdfs}) or logarithmic densities (shown in Figure~\ref{fig:predLogDens}).
    One can see that heavier-tailed heterogeneity priors also yield correspondingly heavier-tailed MAP-priors.
    The heterogeneity and the corresponding MAP priors' properties are also summarized and compared in Table~\ref{tab:mapExamples} in terms of prior quantiles and effective sample sizes~($\elir$).
  \begin{table*}[ht]
  \centering
  \caption{\label{tab:mapExamples}Summaries of MAP priors resulting from several settings for the heterogeneity ($\tau$) prior. The half-normal(0.5) prior is contrasted with half-normal priors of differing scale, as well as with priors of differing distributional families, but with matching prior medians. Note that in the context of the present example, the MAP prior's domain corresponds to logarithmic hazard ratios (log-HRs). Quantiles are centered at~$y_1$}
  \begin{tabular}{lccccrrr}
    \toprule
    \multicolumn{3}{c}{$\tau$ prior} & & standard & \multicolumn{3}{c}{quantile ($\theta_2 - y_1$)} \\
    \cmidrule(lr){1-3} \cmidrule(lr){6-8}
    family                & scale & median & $\elir$ & deviation & 95.0\% & 97.5\% & 99.5\% \\ 
    \midrule
    half-normal           & 0.50 & 0.34 & 26.6 & 0.84 & 1.32 & 1.72 & 2.72 \\[1ex] 
    half-normal           & 0.25 & 0.17 & 45.7 & 0.57 & 0.93 & 1.13 & 1.62 \\
    half-normal           & 1.00 & 0.67 & 12.8 & 1.48 & 2.35 & 3.18 & 5.19 \\[1ex] 
    half-$t_{\nu=4}$      & 0.46 & 0.34 & 25.3 & 1.02 & 1.45 & 1.98 & 3.58 \\ 
    half-Cauchy           & 0.34 & 0.34 & 23.4 &      & 2.45 & 4.85 & 24.02 \\ 
    half-logistic         & 0.31 & 0.34 & 25.8 & 0.91 & 1.39 & 1.85 & 3.09 \\ 
    exponential           & 0.49 & 0.34 & 24.5 & 1.07 & 1.56 & 2.19 & 3.96 \\ 
    Lomax($\alpha\!=\!6$) & 2.75 & 0.34 & 24.0 & 1.31 & 1.70 & 2.50 & 5.05 \\ 
    Lomax($\alpha\!=\!1$) & 0.34 & 0.34 & 23.1 &      & 3.29 & 7.05 & 37.17 \\ 
    \bottomrule
  \end{tabular}
  \end{table*}
  
    It is sometimes also instructive to observe the effects of variations of the prior on the resulting estimates; e.g., varying the heterogeneity prior scale allows for a sensitivity  (or tipping point) analysis. Such an analysis is shown in Appendix~\ref{sec:AlportSensi}; the amount of borrowing in reflected in the shrinkage interval's width, but in the present example inference would not change qualitatively, and a log-HR of zero always remains included.

  \subsection{Heart failure example}\label{sec:cardio}
  The \textsc{Spirit-HF} trial has been designed in order to test the efficacy of spironolactone in patients with heart failure (HF) \cite{SpiritHF}.
  Spironolactone is expected to reduce cardiovascular mortality as well as hospitalizations due to heart failure, and had previously been investigated in the \textsc{Topcat} trial \cite{PittEtAl2014}.
  Both studies refer to the composite of (recurrent) HF~hospitalization and cardiovascular death as the primary endpoint to evaluate treatment efficacy.
  Despite a sizeable sample size of 3445~patients and a mean follow-up duration of more than $3$~years, the \textsc{Topcat} trial failed to demonstrate statistical significance; 
  the estimated hazard ratio (HR) was at 0.89 (0.77, 1.04) ($p=0.14$) \cite{PittEtAl2014}.
  
  The analysis of the new \textsc{Spirit-HF} trial meanwhile is being planned, and may take into consideration the evidence already generated in the \textsc{Topcat} trial. One idea may be to derive a \emph{shrinkage estimate}, anticipating some between-study heterogeneity, and dynamically borrowing information from the earlier study based on the corresponding MAP prior \cite{RoeverFriede2020,RoeverFriede2021}.
  For the between-study heterogeneity~$\tau$, use of a $\halfnormaldistn(0.25)$~prior may be appropriate.
  The heterogeneity prior may be motivated referring to anticipated levels of heterogeneity based on general considerations \cite{RoeverEtAl2021} or using empirical evidence, in particular in view of the similar study designs and the effect measure being a log-HR \cite{LilienthalEtAl2024}.

\begin{figure}[b]
    \centering
    \makebox{\includegraphics[width=0.5\linewidth]{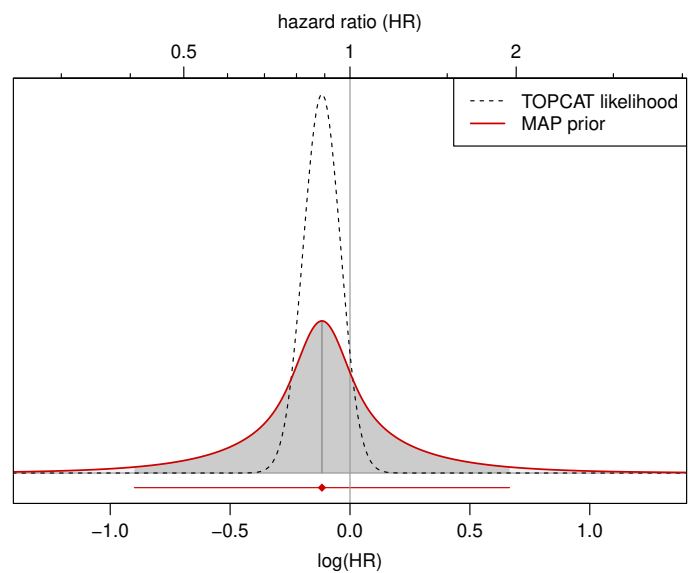}}
    \caption{\label{fig:topcat}Illustration of likelihood and corresponding MAP-prior for the heart failure example, using a $\halfnormaldistn(0.25)$ prior for~$\tau$. The horizontal line at the bottom indicates the 95\% prediction interval} 
  \end{figure}

  In the present case, analysis is based on the loga\-rit\-hmic~HR; the HR estimated in the \textsc{Topcat} trial corresponds to a log-HR of~$-0.117$ with a standard error of~$0.077$. The corresponding \textsc{Topcat} likelihood along with the resulting MAP~prior is illustrated in Figure~\ref{fig:topcat}.
  The variance (squared standard error) of the \textsc{Topcat} study's estimate was~$s_1^2=0.077^2$ while for the assumed heterogeneity prior the expected heterogeneity variance is~$\expect[\tau^2]=0.25^2$ (see also Table~\ref{tab:tauSquared}), so that the resulting MAP~prior's variance~(\ref{eqn:mapVar}) is $s_1^2 + 2\,\expect[\tau^2] = 0.362^2$, and the majority of the variance is due to epistemic uncertainty relating to the anticipated similarity of the \textsc{Topcat} and \textsc{Spirit-HF}
  parameters.
  The 95\% prediction interval for the MAP prior is centered at the \textsc{Topcat} log-HR~estimate and ranges from~$-0.899$ to $+0.665$, corresponding to HRs in the range~[$0.407$, $1.945$].
  According to the MAP prior, the probability of a beneficial treatment effect (a~log-HR below zero) is~71\%.
  The MAP prior has an $\elir$ of~399, i.e, only 12\% of the 3445~actual \textsc{Topcat} patients, and 31\% of the estimated enrolment of 1300~\textsc{Spirit-HF} patients \cite{SpiritHF}.
  This means that the prior derived from the \textsc{Topcat} data will not enter the eventual analysis as an additional 3445~patients (as would be the case if both study populations were pooled na\"{i}vely), but instead we expect an accuracy corresponding to a total of some $1300+399$~patients.
  The \textsc{Spirit-HF} study's contribution to its own shrinkage estimate may also be assessed \cite{RoeverFriede2021}; assuming that both studies show the same dependence of standard error and sample size, the \textsc{Spirit-HF} study will account for a minimum of~$61\%$ in weight to the eventual effect estimate.


\section{Discussion}\label{sec:discussion}
  Despite the seemingly odd notion of a \emph{meta-analysis of a single study}, the use of MAP priors remains completely consistent down to the extreme case of only one data point.
  The ``usual'' toolbox remains available, including common prior specifications \cite{RoeverEtAl2021}, computation of effective sample sizes (ESS) \cite{NeuenschwanderEtAl2020} robustification \cite{SchmidliEtAl2014},
  as well as common meta-analysis software (e.g., the \texttt{bayesmeta} or \texttt{RBesT} \textsf{R}~packages) \cite{Roever2020,WeberEtAl2021} for practical implementation.
  In addition, for $k=1$, there are connections to \emph{bias allowance} and \emph{power prior} models (see Section~\ref{sec:shrink}) 
  that may help help motivating a MAP~approach (or vice versa).
  MAP priors based on few estimates are generally rather heavy-tailed, which will ensure robust operating characteristics \cite{OHaganPericchi2012,RoeverFriede2020,RoeverFriede2021}.
  For few data points in general, and in particular for only a single data point, the prior specification for the heterogeneity parameter~$\tau$ gains in importance and needs to be particularly well-founded and convincing \cite{RoeverEtAl2021}.

  While only the normal model (NNHM) was discussed here, the idea also extends to other model families; for example, derivation of a MAP prior would also work for a binomial-normal model (as implemented in the \texttt{RBesT}~package) \cite{WeberEtAl2021}.
  Another related approach (with some similarity to the power prior) is given by the \emph{commensurate prior} \cite{HobbsEtAl2011}, which, however does not constitute a special case of the MAP~prior.
  Empirical MAP~priors may then be utilized in different ways, either to simply motivate a reasonable sample size (or other design aspects), or to implement explicit borrowing of historical information \cite{SchmidliNeuenschwanderFriede2017,MuehlemannEtAl2023}.
  When MAP~priors are used to also inform the analysis, it is important to approach the evaluation of operating characteristics from a sensible angle; the naive application of classically ``frequentist'' measures to judge a Bayesian procedure, in particular when informative priors are involved, will often not provide a meaningful assessment of its actual features \cite{GneitingEtAl2007,CookGelmanRubin2006,BestEtAl2024}.

  A common concern in the context of the use of historical data is that an informative prior might unduly dominate the eventual analysis; for example, in the heart failure example application, one might be worried that the much larger \textsc{Topcat} trial would swamp the data from the smaller \textsc{Spirit-HF} study. However, for the shrinkage estimate of interest here, the second study's contribution is bounded by a minimum of~$61\%$ within the suggested setup. This proportion would increase for a more conservative heterogeneity prior specification \cite{RoeverEtAl2021} or when implementing robustification \cite{SchmidliEtAl2014}, however, such modelling decisions should probably rather be based on considerations of prior information than on deduced operating characteristics.

  Besides considerations of the value of ``borrowed'' information for a given parameter estimate, MAP~priors based on historical data may also be interesting for the \emph{design} of subsequent trials, with or without the eventual use of shrinkage estimation in the final analysis. Historical information may then help determining sensible ranges for nuisance parameters \cite{SchmidliNeuenschwanderFriede2017} or sample sizes \cite{Lindley1997,BruttiEtAl2014}, for interim decisions \cite{SchmidliEtAl2014,NeuenschwanderEtAl2016b}, or it may be used in a more comprehensive fashion to to ensure a positive joint outcome \cite{NeuenschwanderEtAl2016b,PawelEtAl2023}.

  Van~Zwet \emph{et~al.} (2024) argue that also the analysis of a single study should account for heterogeneity of the treatment effect across studies. Therefore, they propose to consider analyses of individual studies also within an overarching NNHM framework similar to our approach presented here; using informative, empirically motivated priors for both~$\mu$ and~$\tau$, inference may then be focused on the overall mean effect~($\mu$) rather than the study-specific~$\theta_1$ even in the analysis of only a single study \cite{Gelman2024,vanZwetEtAl2025}.






\clearpage
\begin{appendix}
\renewcommand\thefigure{A\arabic{figure}}
\renewcommand\thetable{A\arabic{table}}
\section{Appendix}
\subsection{Posterior predictive distribution}\label{sec:PostPredAppendix}
  When an informative $\normaldistn(\mup, \sigmap^2)$~prior is assumed for the overall mean~$\mu$, the posterior predictive distribution for a ``new'' study-specific mean~$\theta_{k+1}$ (conditional on a given heterogeneity value) is again normal with moments
  \begin{eqnarray}
    \expect[\theta_{k+1}|y_1, \ldots, y_k, s_1, \ldots, s_k, \tau] 
    & = & \frac{\frac{\mup}{\sigmap^2}+\sum_{i=1}^k \frac{y_i}{s_i^2+\tau^2}}{\frac{1}{\sigmap^2}+\sum_{i=1}^k\frac{1}{s_i^2+\tau^2}}
    \\
    \var(\theta_{k+1}|y_1, \ldots, y_k, s_1, \ldots, s_k, \tau) 
    & = & \frac{1}{\frac{1}{\sigmap^2}+\sum_{i=1}^k\frac{1}{s_i^2+\tau^2}} + \tau^2
  \end{eqnarray}
  implying for the specific case of $k=1$ that
  \begin{eqnarray}
    \expect[\theta_{2}|y_1, s_1, \tau] 
    & = & \mup\frac{s_1^2+\tau^2}{\sigmap^2+s_1^2+\tau^2}
            + y_1 \frac{\sigmap^2}{\sigmap^2+s_1^2+\tau^2}
    \\
    \var(\theta_{2}|y_1, s_1, \tau) 
    & = & \frac{1}{\frac{1}{\sigmap^2} + \frac{1}{s_1^2 + \tau^2}} + \tau^2
  \end{eqnarray}
  (see R\"{o}ver~(2020) \cite{Roever2020}).
  One can already see that in the limiting case of an increasingly vague effect prior ($\sigmap\rightarrow\infty)$, the prior's influence vanishes, and the (conditional) variance increases.
  
  Specification of an improper uniform prior for the overall mean effect~$\mu$ also leads to a proper posterior; the posterior predictive moments then are of the slightly simpler form
  \begin{eqnarray}
    \expect[\theta_{k+1}|y_1, \ldots, y_k, s_1, \ldots, s_k, \tau] 
    & = & \frac{\sum_{i=1}^k \frac{y_i}{s_i^2+\tau^2}}{\sum_{i=1}^k\frac{1}{s_i^2+\tau^2}}
    \\
    \var(\theta_{k+1}|y_1, \ldots, y_k, s_1, \ldots, s_k, \tau) 
    & = & \frac{1}{\sum_{i=1}^k\frac{1}{s_i^2+\tau^2}} + \tau^2
  \end{eqnarray}
  (see R\"{o}ver~(2020) \cite{Roever2020}).
  In the case of a single study ($k=1$), these expressions then simplify to
  \begin{equation}\label{eqn:ImproperPredMoments}
    \expect[\theta_{2}|y_1, s_1, \tau] 
    \; = \; y_1
    \quad \mbox{and} \quad
    \var(\theta_{2}|y_1, s_1, \tau) 
    \; = \; s_1^2 + 2\tau^2
  \end{equation}

\subsection{Gain in effective sample size}\label{sec:EssGainAppendix}
  The \emph{relative} gain in information from a prior may be quantified using the \emph{gain in effective sample size}, which is based on the relative width of 95\%~intervals \emph{with} and \emph{without} consideration of the (informative) prior. 
  First, consider the relative width~$q$, the ratio of interval widths (or standard errors) ($q=\frac{\mbox{\small width using informative prior}}{\mbox{\small width using vague prior}}$).
  Assuming that standard errors are proportional to the inverse of the square root of the sample size, the gain in effective sample size then is given by~$q^{-2}-1$. For example, if the informative prior yields an interval that is only half as wide ($q=0.5$), this would otherwise have required a quadrupled sample size, or a $q^{-2}-1=300\%$ increase. If the interval is 90\% as wide ($q=0.9$), this corresponds to an approximate $q^{-2}-1=23\%$ increase \cite{RoeverFriede2020}.

\subsection{Heterogeneity priors}\label{sec:PriorAppendix}
\begin{figure}[b]
    \centering
    \makebox{\includegraphics[width=0.5\linewidth]{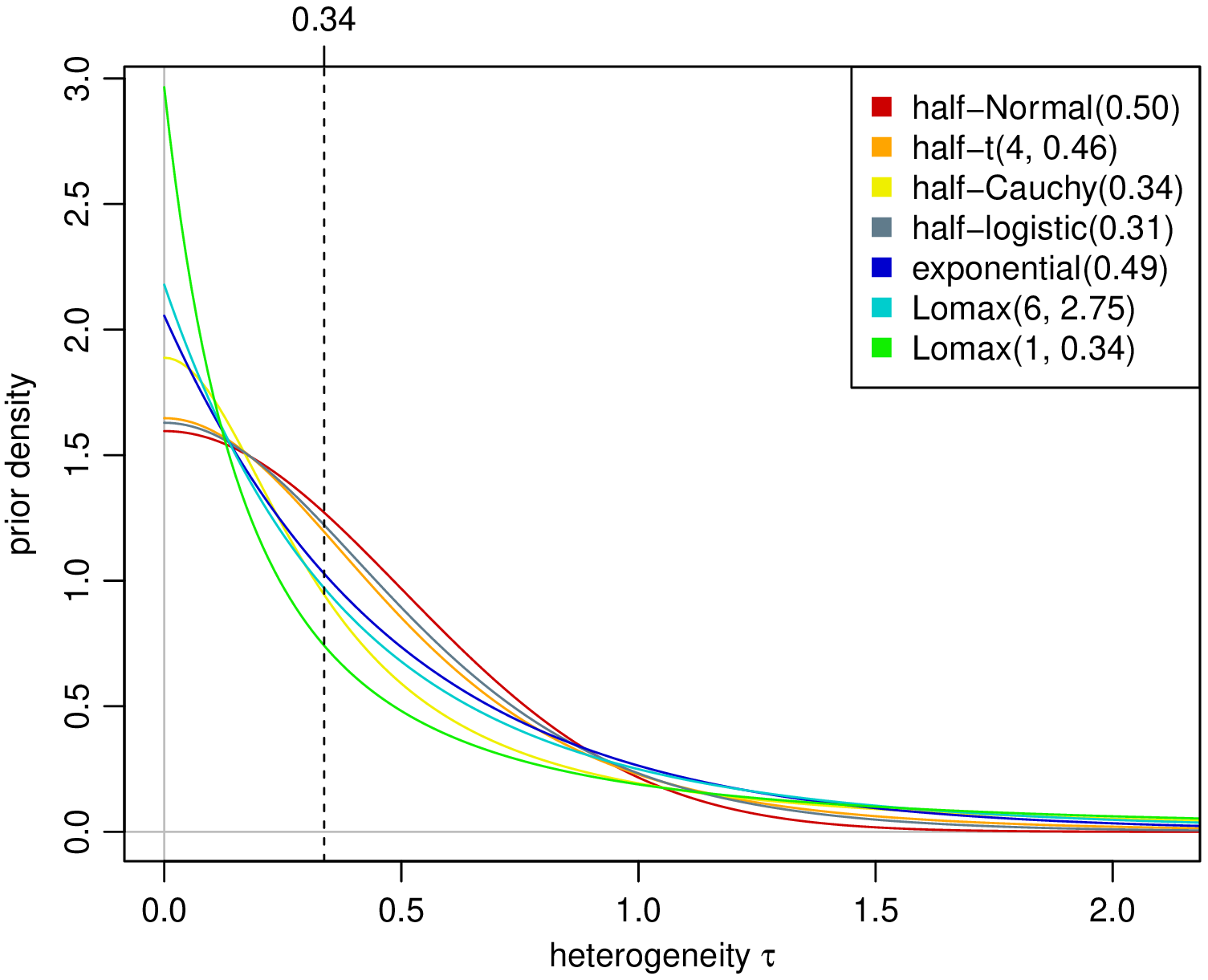}}
    \caption{\label{fig:tauPriorDensities}Illustration of the several heterogeneity priors compared in Section~\ref{sec:variations} in terms of their probability density functions. All priors are scaled such that they have a common median of 0.34 (the median of a half-normal(0.5) prior; dashed line)}
  \end{figure}

  Figure~\ref{fig:tauPriorDensities} shows several prior densities discussed in Section~\ref{sec:variations} that are all scaled to a common median (of~0.34, the median of a HN(0.5) distribution).
  
  \begin{figure}
    \centering
    \makebox{\includegraphics[width=0.5\linewidth]{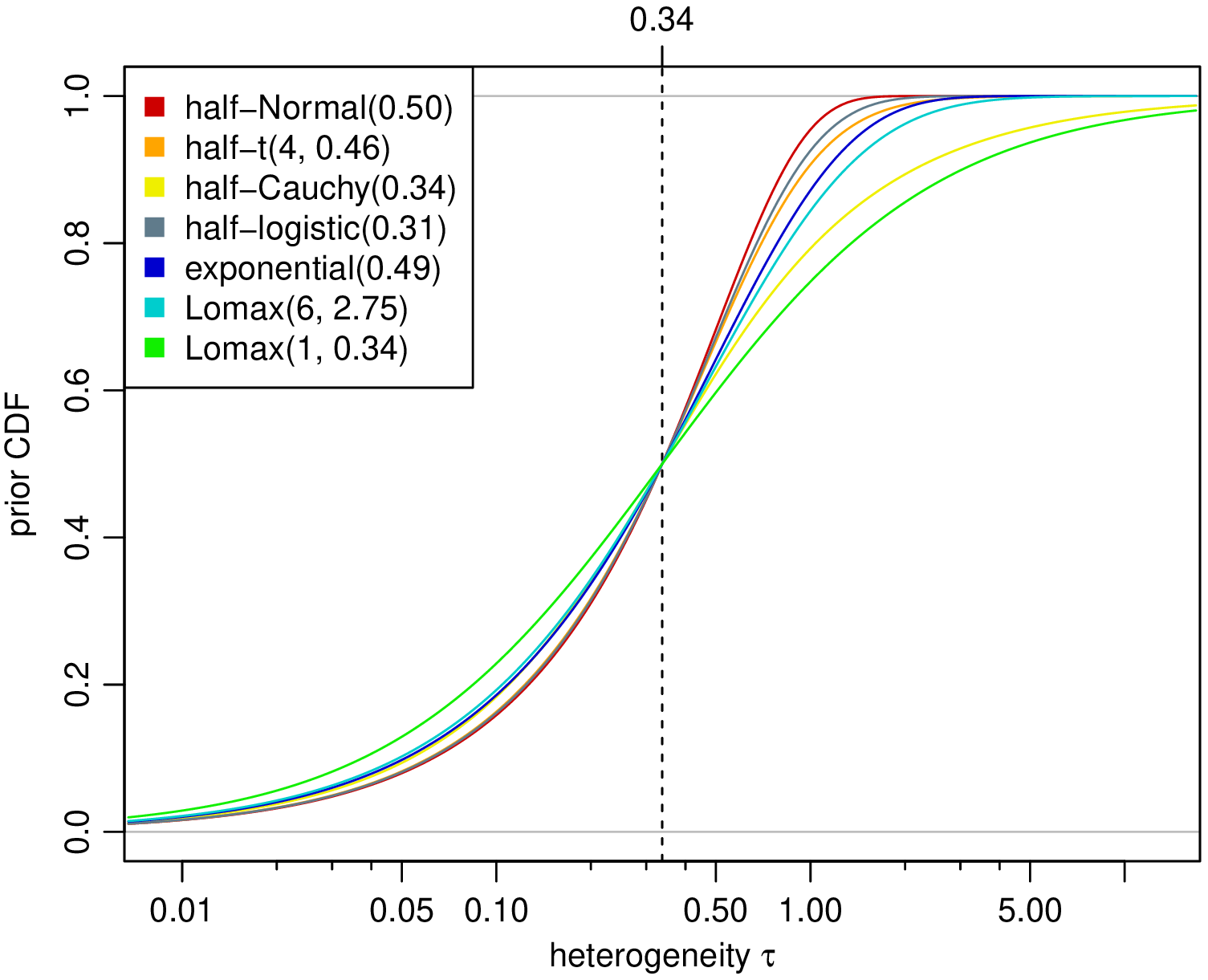}}
    \caption{\label{fig:tauPriorCDFs}Illustration of the several heterogeneity priors compared in Section~\ref{sec:variations} in terms of their cumulative distribution functions (CDFs). All priors are scaled such that they have a common median of 0.34 (dashed line)}
  \end{figure}

  \subsection{Variance of the MAP prior}\label{sec:MapVarAppendix}
    The prior predictive distribution $p(\theta_2|y_1,s_1)$ is a normal \emph{scale mixture} with (constant) mean~$y_1$ and (conditional) variance~$s_1^2+2\tau^2$, where $\tau$ is distributed according to the specified heterogeneity prior.
    The mixture distribution's marginal variance results 
    as $\var(\theta_2|y_1,s_1) = s_1^2 + 2\,\expect[\tau^2]$ and depends on the (prior) expectation of the \emph{squared} heterogeneity~$\expect[\tau^2]$ (see~(\ref{eqn:mapVar})).
    \begin{table*}[t]
      \caption{\label{tab:tauSquared}Expected values of~$\tau^2$ based on various common (prior) distributions for~$\tau$, depending on their scale parameter~$s$. An asterisk~($\ast$) indicates that there is no simple analytical expression. The half-Cauchy prior would be an additional option, but does not have a finite expectation (it is in fact also a special case of the half-Student\mbox{-}$t$ prior, with $\nu=1$ degree of freedom)}
      \centering
      {\footnotesize
      \begin{tabular}{lccclcl}
        \toprule
          \multicolumn{1}{c}{} & \multicolumn{4}{c}{standard deviation~$\tau$} & \multicolumn{2}{c}{variance~$\tau^2$} \\
        \cmidrule(lr){2-5}
        \cmidrule(lr){6-7}
          \multicolumn{1}{l}{distribution} & median & 95\% quantile & expectation & & expectation & \\
        \midrule
         $\tau\sim\halfnormaldistn(s)$   & $\approx 0.674s$ & $\approx 1.96s$ & $\sqrt{\frac{2}{\pi}}\,s \approx 0.798s$ & & $s^2$ & \\
         $\tau\sim\halftdistn_\nu(s)$    & $\ast$ & $\ast$ & $2\sqrt{\frac{\nu}{\pi}}\frac{\Gamma(\frac{\nu+1}{2})}{\Gamma(\frac{\nu}{2})(\nu-1)}s$ & $(\nu>1) $ & $\frac{\nu}{\nu-2}s^2$ & $(\nu>2)$\\
         $\tau\sim\halflogidistn(s)$     & $\log(3)\,s \approx 1.10s$ & $\approx 3.66s$ & $\log(4)\,s \approx 1.39s$ & & $\frac{\pi^2}{3}s^2\approx 3.29s^2$ & \\
         $\tau\sim\expodistn(s)$         & $\log(2)\,s \approx 0.693s$ & $\approx 3.00s$ & $s$ & & $2s^2$ & \\
         $\tau\sim\lomaxdistn(\alpha,s)$ & $(2^\frac{1}{\alpha}-1)\,s$ & $(20^\frac{1}{\alpha}-1)\,s$ & $\frac{1}{\alpha-1}s$ & $(\alpha>1)$ & $2\frac{\Gamma(\alpha-2)}{\Gamma(\alpha)}s^2$ & $(\alpha>2)$\\
         $\tau\sim\unifdistn[0,s]$ & $\frac{s}{2}$ & $0.95s$ & $\frac{s}{2}$ & & $\frac{1}{3}s^2$ & \\
        \bottomrule
      \end{tabular}
      }
    \end{table*}
    Table~\ref{tab:tauSquared} below summarizes expectations for~$\tau^2$ for a range of common heterogeneity priors.
    Note also the related Table~B1 in the appendix of R\"{o}ver \emph{et~al.} (2021) \cite{RoeverEtAl2021} giving some additional details for the prior distributions shown here.
    
  \subsection{Power prior exponent's distribution}\label{sec:PowerPriorExponent}
\begin{figure}
      \centering
      \makebox{\includegraphics[width=0.5\linewidth]{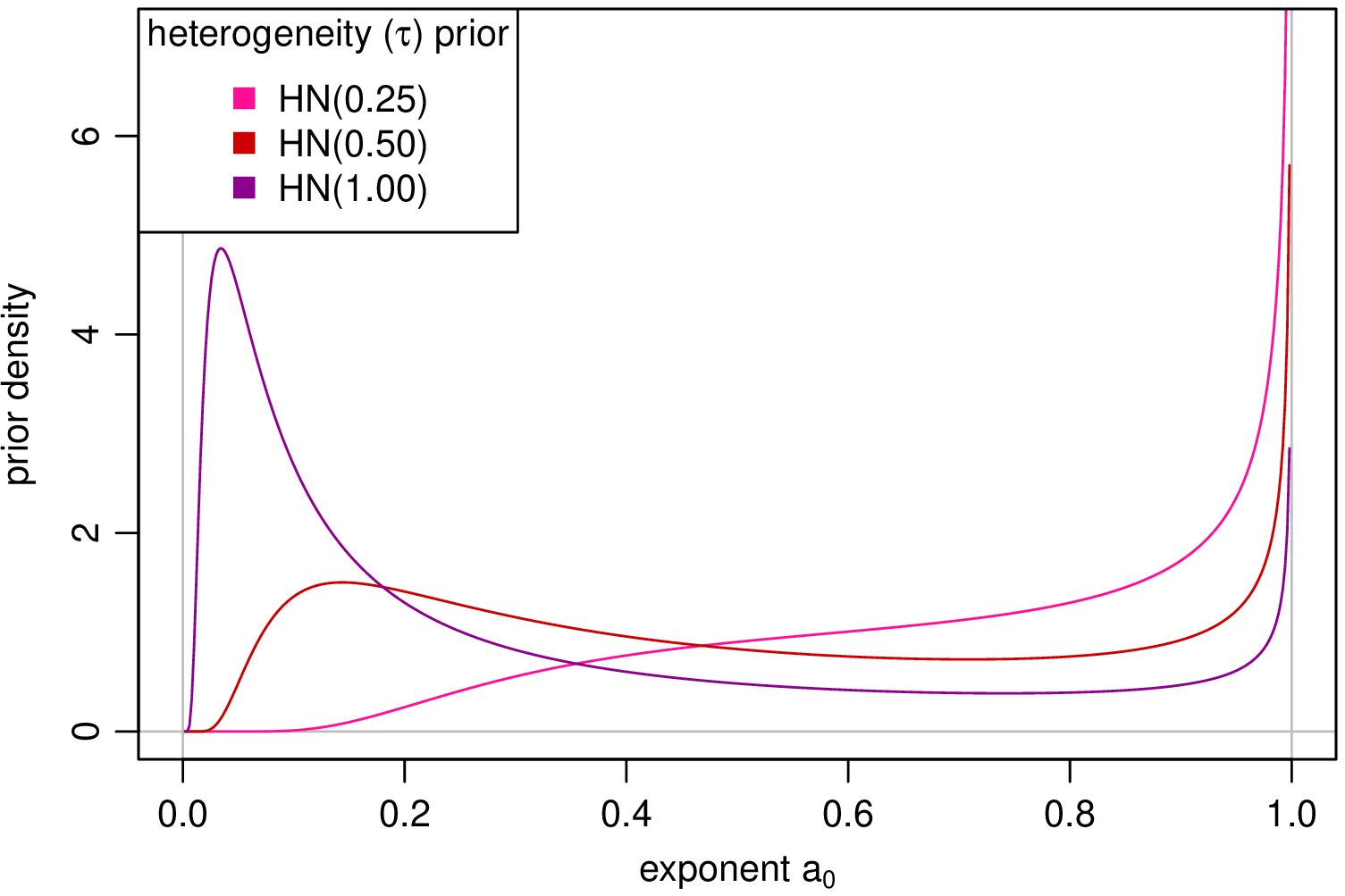}}
      \caption{\label{fig:powerDensities}Illustration of prior distributions for the power prior exponent~$a_0$ corresponding to certain prior distributions assumed for the heterogeneity~$\tau$ (and $s_1=0.451$)}
    \end{figure}

    For any \emph{fixed} heterogeneity value~$\tau$, the (conditional) MAP prior is equivalent to a power prior with exponent~$a_0=\bigl(2\frac{\tau^2}{s_1^2}+1\bigr)^{-1}$ (see also Section~\ref{sec:powerPriorConnection}) \cite{RoeverFriede2020}.
    Through this functional relationship, any prior density~$p_\star(\tau)$ for the heterogeneity implies a corresponding prior for the exponent~$a_0$ with probability density function
    \begin{equation}\textstyle
      p(a_0) \;=\; 
      \frac{s_1}{2\sqrt{2}}
      \frac{\sqrt{\frac{a_0}{1-a_0}}}{a^2}\;
      p_\star\biggl(s_1\sqrt{\frac{1-a_0}{2\,a_0}}\biggr) \mbox{.}
    \end{equation}
    Figure~\ref{fig:powerDensities} illustrates such densities for an example value of $s_1=0.451$ (as in the Alport example from Section~\ref{sec:alport}).
    The prior densities for~$a_0$ are shown for half-normal priors with scales~0.25, 0.50 and~1.00 (corresponding to the cases also illustrated in Figure~\ref{fig:scaleVariations}).
    
    A value of~$a_0\!=\!1$ for the exponent corresponds to \emph{full borrowing}, while smaller values imply increasing degrees of discounting of prior information. While the $\halfnormaldistn(0.25)$ prior places a substantial share of prior probability near~\mbox{$a_0\!=\!1$}, larger prior scale parameters correspond to less a-priori expected borrowing, eventually resulting in bimodal priors for~$a_0$.

    Note that (as elaborated in Section~\ref{sec:powerPriorConnection}) the mapping between~$\tau$ ans~$a_0$ always depends on the ``source'' study's standard error~($s_1$). A prior for~$\tau$ may be motivated independent of~$s_1$, while specification of a value (or distribution) for~$a_0$ may have odd consequences when varying the source study's size or precision~($s_1$).

  \subsection{Sensitivity analysis (Alport example)}\label{sec:AlportSensi}
    In the Alport example application from Section~\ref{sec:alport}, varying the half-Normal heterogeneity prior's scale parameter affects the precision of the corresponding MAP~prior, and with that, the eventual amount of borrowing from the observational data.
    \begin{figure}
      \centering
      \makebox{\includegraphics[width=0.5\linewidth]{figures/alport-sensitivity.eps}}
      \caption{\label{fig:AlportSensi}Illustration of the effect of varying the (half-normal) heterogeneity prior's scale on the resulting RCT~shrinkage estimate from Section~\ref{sec:alport}}
    \end{figure}
    Figure~\ref{fig:AlportSensi} shows how the resulting shrinkage estimate for the RCT is affected. A smaller prior scale leads to more borrowing and hence a shorter shrinkage interval eventually approaching the common-effect estimate. A larger prior scale leads to less borrowing, with the shrinkage interval eventually approaching the interval based on the RCT~data alone. Since a log-HR of zero always remains included, in this example there is no ``tipping point'' for the heterogeneity's prior scale parameter.
    

  \subsection{Example \textsf{R}~code}
  \subsubsection{Alport example}
{
\begin{verbatim}
# data from:
#   Gross et al. (2020)
#   https://doi.org/10.1016/j.kint.2019.12.015
#   Figure 2(b)

GrossEtAl2020 <- cbind.data.frame("data"     = c("observational", "RCT"),
                                  "patients" = c(70, 20),
                                  "events"   = c(29, 8),
                                  "hr"       = c(0.53, 0.51),
                                  "lower"    = c(0.22, 0.12),
                                  "upper"    = c(1.29, 2.20))
GrossEtAl2020$loghr <- log(GrossEtAl2020$hr)
GrossEtAl2020$se    <- (log(GrossEtAl2020$upper)-log(GrossEtAl2020$lower))/(2*qnorm(0.975))

# show data (see Table 1):
GrossEtAl2020

# convert effect measures to "escalc" object:
library("metafor")
es <- escalc(measure="GEN",
             yi=loghr, sei=se, ni=patients,
             slab=data, data=GrossEtAl2020)


###############################################################
# perform joint meta-analysis of both estimates (MAC approach):
library("bayesmeta")
bm <- bayesmeta(es,
                tau.prior=function(t){dhalfnormal(t, scale=0.50)})

# show data as well as overall and shrinkage estimates:
forestplot(bm, predict=FALSE,
           xlab="log-HR",
           txt_gp = fpTxtGp(ticks = gpar(cex=1), xlab = gpar(cex=1)))

# check out shrinkage estimate(s):
bm$theta  # (log-HRs)
exp(bm$theta[c("median","95% lower", "95% upper"),"RCT"])  # (HR)


##########################################################
# meta analysis of observational data only (MAP approach):
map <- bayesmeta(es[1,],
                 tau.prior=function(t){dhalfnormal(t, scale=0.50)})

# check out predictive distribution (MAP prior):
map$summary[,"theta"]  # (log-HR)


###############################################################
# illustrate prior, likelihood, posterior / shrinkage estimate
# (see Figure 1):

colvec <- c("map"    = "red3",
            "likeli" = "blue",
            "shrink" = "green3")
x <- seq(-4, 2.5, length=200)
plot(c(-3.5, 2.0), c(0.0, 0.85), type="n",
     main="Alport example (Gross et al., 2020)",
     xlab="log-HR", ylab="")
# MAP prior (predictive density):
lines(x, map$dposterior(theta=x, predict=TRUE), type="l", col=colvec[1])
# RCT likelihood:
lines(x, dnorm(x, mean=es$yi[2], sd=sqrt(es$vi[2])), type="l", col=colvec[2])
# posterior (=shrinkage estimate):
lines(x, bm$dposterior(theta=x, individual="RCT"), type="l", col=colvec[3])
abline(h=0, col="grey20")
lines(c(0,0), c(0, 1), col="grey20", lty=2)
legend("topleft", c("MAP prior (observational)", "likelihood (RCT)",
                    "shrinkage (RCT)"),
       col=colvec, pch=15)
\end{verbatim}
}

  \subsubsection{Heart failure example}
{
\begin{verbatim}
# TOPCAT study (Pitt et al., 2024) quotes a HR of:
# 0.89 (0.77, 1.04)

# log-HR:
log(0.89)  # = -0.117

# standard error:
(log(1.04) - log(0.77)) / (2*qnorm(0.975))  # = 0.077

# derive MAP prior:
library("bayesmeta")
bm <- bayesmeta(y=-0.117, sigma=0.077, label="Topcat",
                tau.prior=function(t){dhalfnormal(t, scale=0.25)})

# (MAP-) prior predictive interval:
round(rbind("log-HR"=bm$summary[c(2,5:6), "theta"],
            "HR"=exp(bm$summary[c(2,5:6), "theta"])), 3)
# prior probability of HR<0:
bm$pposterior(theta=0, predict=TRUE)

###########################################
# illustrate TOPCAT likelihood
# and associated MAP prior (see Figure 6):
x <- seq(-1.8, 1.8, length=200)
plot(c(-1.3, 1.3), c(0.0, 5.18), type="n", 
     xlab="log-HR", ylab="")

# TOPCAT likelihood:
lines(x, dnorm(x, mean=-0.117, sd=0.077), lty=2)
# MAP prior:
lines(x, bm$dposterior(theta=x, predict=TRUE), col="red3", lwd=1.5)
# axes, legend:
abline(h=0, v=0, col="grey60")
legend("topright", lty=c(2,1), col=c("black", "red3"), lwd=c(1,1.5),
       c("TOPCAT likelihood", "MAP prior"))


# compute TOPCAT's unit information standard deviation (UISD):
# (standard error was 0.077, based on 3445 patients)
uisd(n=3445, sigma=0.077)  # = 4.5

# compute MAP prior's effective sample size (ESS_ELIR):
ess(bm, uisd=4.5)  # = 399 patients
\end{verbatim}
}
\end{appendix}




\paragraph{Funding Statement}
  Support from the German Centre for Cardiovascular Research \emph{(Deutsches Zentrum für Herz-Kreislauf-Forschung e.V., DZHK)} is gratefully acknowledged (grant number \mbox{81Z0300108}).

\paragraph{Competing Interests}
  The authors have declared no conflict of interest.

\paragraph{Data Availability Statement}
  The data and code that supports the findings of this study are available in the supplemental material of this article.


\paragraph{Author Contributions}
Conceptualization: T.F.; C.R. Methodology: C.R. 
Writing original draft: C.R. All authors approved the final submitted draft.


\bibliographystyle{wileyNJD-AMA}
\bibliography{literature}


\end{document}